\documentclass[reprint,amsmath,amssymb,aps,floats,superscriptaddress,preprintnumbers]{revtex4-2}

\usepackage{graphicx}
\usepackage{dcolumn}
\usepackage{bm}
\usepackage{hyperref}

\usepackage{color}
\usepackage{cancel}
\usepackage[normalem]{ulem}
\usepackage{AAS_macro}
\usepackage{multirow}
\usepackage{orcidlink}

\usepackage{amsmath}	
\usepackage{amssymb}	
\usepackage{mathtools}

\renewcommand{\arraystretch}{1.4}

\begin{document}

\title{Observational constraint on Dark Energy from Quantum Uncertainty}

\author{Long Huang}
\affiliation{Xinjiang Astronomical Observatory, Chinese Academy of Sciences, Urumqi 830011, China.}
\affiliation{University of Chinese Academy of Sciences, Beijing, 10039, China.}

\author{ Xiaofeng Yang}
\affiliation{Xinjiang Astronomical Observatory, Chinese Academy of Sciences, Urumqi 830011, China.}
\affiliation{Key Laboratory of Radio Astronomy, Chinese Academy of Sciences, Nanjing 210008, China.}
\affiliation{Key Laboratory of Radio Astrophysics in Xinjiang Province, Urumqi 830011, China.}

\author{ Xiang Liu}
\affiliation{Xinjiang Astronomical Observatory, Chinese Academy of Sciences, Urumqi 830011, China.}
\affiliation{Key Laboratory of Radio Astronomy, Chinese Academy of Sciences, Nanjing 210008, China.}
\affiliation{Key Laboratory of Radio Astrophysics in Xinjiang Province, Urumqi 830011, China.}

\begin{abstract}
 We explore the theoretical possibility that dark energy density is derived from the vacuum particle pairs together with the quantum fluctuation of space-time. By assuming the vacuum particle pairs fall into the horizon boundary of the cosmos with the expansion of the universe, we can deduce the uncertainty in the relative position of particle pairs based on the quantum fluctuation of space-time, which can be used to estimate their or dark energy density. Furthermore, we attempt to explain the origin of negative pressure from the increasing entropy density of the Boltzmann system and derive the equation for the state parameter, which is consistent with the specific equations of state for dark energy. Finally, we fit the models to the SNIa Pantheon sample and Planck 2018 CMB angular power spectra and give statistical results for the cosmology parameters.
 \end{abstract}

\maketitle

\section{Introduction}

Since 1998, when Riess et al. used SNIa data to statistically indicate that the expansion of the universe is accelerating \cite{riess1998},  physicists have been providing various theories to explain this acceleration, including $f(R)$ theory \cite{Capo}, Brans-Dicke theory \cite{Brans}, and dark energy theory \cite{Peebles}. At present,  the dark energy theory can be used to effectively explain the cosmic microwave background (CMB) anisotropies  \cite{Hu}. However, this study mainly focuses on the physical nature of dark energy. Dark energy can be studied using two main approaches. The first is to focus on the properties of dark energy, investigating whether or not its density evolves with time;  this can be verified by reconstructing the equation of state $w(z)$ for dark energy, which is independent of physical models. The reconstruction of the equation of state involves parametric and non-parametric methods \cite{Linder}, the latter including the Principal Component Analysis \cite{Huterer, Clarkson}, Gaussian Processes \cite{Holsclaw, Shafieloo, Seikel}, PCA with the smoothness prior method \cite{Crittenden2009, Crittenden2012, Zhao}, and PCA based on the Ridge Regression Approach\cite{Huang}. The second involves dark energy physical models that are presented from the physical origin of its density and pressure, including scalar field models \, cite{Ratra}, pseudo-Nambu-Goldstone bosons for cosmology \cite{Frieman}, holographic dark energy \cite{Li}, age dark energy \cite{Mazia2007} and so on.

Currently, it may be difficult to judge which model, method, or result is more persuasive, however,  a model of dark energy that concerns its physical nature is essential. From the point of view of the models, Maziashvili presented a method that uses the Krolyhazy relation and time-energy uncertainty relation to estimate the density of dark energy  \cite{Mazia2007, Mazia20076}, and the result is consistent with astronomical data if the unique numerical parameter in the dark energy model is taken to be on the order of one.

Based on this, to further explore the origin of dark energy density and pressure, We explore the theoretical possibility that dark energy density is derived from vacuum particle pairs together with the quantum fluctuation of space-time.  We can deduce the uncertainty in the position of particle pairs based on the quantum fluctuation of space-time, which can be used to estimate the particle pairs and dark energy density.

\section{The quantum fluctuation of space-time and P-symmetry }

The quantum fluctuation of space-time relates to the quantum properties of objects. Using the Heisenberg position-momentum uncertainty relations, Wigner derived a quantum limit on the measurability of a certain length \cite{Wigner}. If $t$ is the time required by the measurement procedure, the uncertainty in the length measurement is

\begin{eqnarray}\label{eq1}
\delta t \sim \sqrt {\frac{t}{{{M_c}}}}
\end{eqnarray}
where ${M_c}$ is the mass of the clock, we take the units $\hbar  = c = 1$  throughout this study.

As a result of the above relation, when ${M_c} \to \infty $,  $\delta t \to 0$. To solve this situation, the quantum fluctuations of space-time itself are presented \cite{Karo, Jack, Amelino1994, Amelino1999}. It also results in uncertainty in distance measurements. Thus,  at very short distance scales, space-time is foamy, and the limitation of space-time distance measurements can be given by

\begin{eqnarray}\label{eq3}
\delta t \sim \left\{ \begin{array}{l}
{({t_p}t)^{\frac{1}{2}}}{\kern 1pt} {\kern 1pt} {\kern 1pt} {\kern 1pt} {\kern 1pt} {\kern 1pt} {\kern 1pt} {\kern 1pt} {\kern 1pt} {\kern 1pt} {\kern 1pt} {\kern 1pt} {\kern 1pt} {\kern 1pt} {\kern 1pt} {\kern 1pt} r > {t_p}\\
{t_p}{\kern 1pt} {\kern 1pt} {\kern 1pt} {\kern 1pt} {\kern 1pt} {\kern 1pt} {\kern 1pt} {\kern 1pt} {\kern 1pt} {\kern 1pt} {\kern 1pt} {\kern 1pt} {\kern 1pt} {\kern 1pt} {\kern 1pt} {\kern 1pt} {\kern 1pt} {\kern 1pt} {\kern 1pt} {\kern 1pt} {\kern 1pt} {\kern 1pt} {\kern 1pt} {\kern 1pt} {\kern 1pt} {\kern 1pt} {\kern 1pt} {\kern 1pt} {\kern 1pt} {\kern 1pt} {\kern 1pt} {\kern 1pt} {\kern 1pt} {\kern 1pt} {\kern 1pt} r \le {t_p}
\end{array} \right.
\end{eqnarray}
where ${t_p}$ is the Planck time and ${t_p} = \sqrt {\hbar G} $. This limitation of space-time measurements can be interpreted as the result of quantum fluctuations of space-time. Meanwhile, Eq. (\ref{eq3}) can be derived for massless particles in the framework of $\kappa $-deformed Poincare symmetries \cite{Amelino1997,Amelino1998}.

 Krolyhazy derived another method for describing the quantum fluctuations of space-time, known as  the Krolyhazy relation
\begin{eqnarray}\label{eq4}
\delta t \sim \;{t_p}^{2/3}{t^{1/3}}.
\end{eqnarray}

 We consider that this quantum fluctuation effect is also applicable to the expanding universe.  By assuming massless particle pairs fall into the horizon boundary of the cosmos,  with the expansion of the universe,  the quantum fluctuation of space-time provides particle pairs with nonzero energy.
hence,  the uncertainty in the position of the particle pair is
  \begin{eqnarray}\label{eq6}
\delta {t_{particle}} \sim \delta t.
\end{eqnarray}

 We hypothesize that the mean distance between particle pairs is no less than the uncertainty relation $\delta {t_{scalar}}$, as a result, we obtain the number density of massless particle pairs
\begin{eqnarray}\label{eq7}
N = \delta {t^{ - 3}}.
\end{eqnarray}

If the quantum fluctuation of space-time provides particle pairs with nonzero energy,  we can consider the wave function ${\Psi _0}$  for the massless particle pairs is the non-stationary state. Its wave length can be determined by the uncertainty relation $\delta t$; hence,  ${\Psi _0}$ can be written as the superposition of at least $n$ stationary states ${\varphi _1},{\varphi _2},...,{\varphi _n}$
\begin{eqnarray}\label{eq8}
{\Psi _0} = \sum\limits_k^n {{\alpha _k}} {\varphi _k}{e^{ - i{w_k}t'}},
\end{eqnarray}
where ${{\rm{w}}_k} = 2\pi k/\delta t.$ It is easy to see that the  state components of $\varphi ,{e^{ - i\omega {t'}}}$ are  orthogonal in space and time, respectively.

Alternatively, the wave length of particle
pairs can be determined by the age of universe $t$; hence, the wave function of particle
pairs ${\Psi _0}$ can be written as
\begin{eqnarray}\label{eq9}
{\Psi _0} = \sum\limits_k^n {{{\bar \alpha }_k}} {\bar \varphi _k}{e^{ - i{{\bar w}_k}t'}},
\end{eqnarray}
where ${{\rm{\bar w}}_k} = 2\pi k/t.$ 

\section{ The density and negative pressure for Boltzmann system}
When all particle
pairs are in the ground state ${\varepsilon _0}$, that is
\begin{eqnarray}\label{eq16}
{\varepsilon _0} = \delta {t^{ - 1}}{\kern 1pt} {\kern 1pt} {\kern 1pt} or{\kern 1pt} {\kern 1pt} {\kern 1pt} {t^{ - 1}},
\end{eqnarray}

we can use Eq. (\ref{eq3}) or (\ref{eq4}), (\ref{eq7}) and (\ref{eq16}) to obtain the internal energy per unit volume for massless particle
pairs
\begin{eqnarray}\label{eq17}
{\rm{u = }}{\varepsilon _0}N \sim {({t_p}t)^{ - 2}}.
\end{eqnarray}
Next, we use the increase in entropy to derive the existence of negative pressure.

We consider the Boltzmann system to be composed of massless particle pairs in vacuum; the particles have a quantum state number $k$ for energy level, and it is clear that  the microstate numbers of the Boltzmann system are ${3^N}$; hence, the system has statistical entropy per unit volume
\begin{eqnarray}\label{eq18}
s = {k_B}\ln {k^N},
\end{eqnarray}
where ${k_B}$ is the Boltzmann constant.

 Using the thermodynamic entropy definition $TdS = dU + pdV$,  where $T$ is temperature, and the equations $dS = sdV+Vds$, $dU = udV+Vdu$; with Eq. (\ref{eq3}), (\ref{eq7}), (\ref{eq17}) and (\ref{eq18}), we get
 \begin{eqnarray}\label{eq19}
T = {(\ln {k_c})^{ - 1}}{k_B}^{ - 1}{({t_p}t)^{ - 1/2}}4/3.
\end{eqnarray}
Because $u{({t_p}t)^2}$ is constant,  ${k_c}$ is also constant.

We assume the energy level ${k} $ is increasing gradually with the expanding universe, which can be expressed as
 \begin{eqnarray}\label{eq20}
k:1 \to 2 \to 3 \cdot  \cdot  \cdot  \to {k_c},
\end{eqnarray}

Hence, with the rise in ${k} $, the entropy density will increase, which can generate the negative pressure. From Eq. (\ref{eq17}) and (\ref{eq18}) (\ref{eq20}), it is easy to find that the negative pressure ${p_u} = {\omega _u}u$ satisfies
 \begin{eqnarray}\label{eq21}
{k_B} T\ln {k}^N = (1 + {\omega _u})u,
\end{eqnarray}
solving the above equation yields
 \begin{eqnarray}\label{eq22}
{\omega _u} =  - 1 + \frac{4}{3}\frac{{\ln k}}{{\ln {k_c}}}.
\end{eqnarray}
This is the equation of state for the particle pairs energy in the vacuum. It has an interesting feature. When ${k} \to 1$, one has ${\omega _u} \to  - 1$, and when ${k} \to {k_c}$, ${\omega _u} \to  1/3$, if ${k} \to \sqrt {{k_c}} $, ${\omega _u} \to  - 1/3$. As a result of the above relation, it implies that the vacuum energy is from ordered to disordered state along with time, and it can't violate the principle of entropy increase. Namely, the negative pressure can be derived from the increasing entropy, which avoids the universe keep acceleratedly expanding in the future.

\section{Dark energy density and pressure}
\subsection{dark energy density and the evolution of ground state energy}
We assume that dark energy is composed of massless particle pairs in vacuum, using Eq. (\ref{eq17}), and obtain the dark energy density ${\rho _{de}} \sim {({t_p}t)^{ - 2}}$. We introduce a energy level constant to the dark energy density. Hence, the energy density can be written as
\begin{eqnarray}\label{eq23}
{\rho _{de}} = \frac{{3{n^2}{M_p}^2}}{{{t ^2}}},
\end{eqnarray}
where ${M_p}^{ - 2} = 8\pi G$, ${n^2}$ is an introduced energy level constant and satisfies ${n^2} = k$ in Eq. (\ref{eq8}), $t$ is the age of the universe.

Suppose the universe be spatially flat, define the fraction energy density of dark matter as  ${\Omega _m} = {\rho _m}/3{M_p}^2{H^2}$, and ${\Omega _{de}} = {\rho _{de}}/3{M_p}^2{H^2}$, one has ${\Omega _m} =1- {\Omega _{de}}$ from the Friedmann equation. With the Eq. (\ref{eq23}), ${\Omega _{de}}$ can be written
\begin{eqnarray}\label{eq25}
{\Omega _{de}} = \frac{{{n^2}}}{{{H^2}{t ^2}}} ,
\end{eqnarray}
where $H = \dot a/a$ is the Hubble parameter.
Using the energy conservation equation ${\dot \rho _{de}} + 3H({\rho _{de}} + {p_{de}}) = 0$, with Eq. (\ref{eq23}) (\ref{eq25}), one can get the equation of state of energy density for ${\omega _{de}} = {p_{de}}/{\rho _{de}}$ as
\begin{eqnarray}\label{eq26}
{\omega _{de}} =  - 1 + \frac{2}{{3n}}  .
\end{eqnarray}
At an earlier time, when ${n} \to 1$, ${\omega _u} \to  - 1/3$, and at a later time when $n \to \infty $, ${\omega _u} \to  - 1$.

We also consider to choose the conformal time $\eta $ as the time scale $t$ \cite{Cai,Wei}, and the energy density can be written as
\begin{eqnarray}\label{eqq23}
{\Omega _{de}} = \frac{{{n^2}}}{{{H^2}{\eta ^2}}} ,
\end{eqnarray}
where $\eta $ is the conformal time, which is given by
\begin{eqnarray}\label{eqq24}
\eta  = \int_0^t {\frac{{dt}}{a}} .
\end{eqnarray}
where $a$ is the scale factor of our universe. We take the present scale factor ${a_0} = 1$. 

Then combining with the dark energy density conservation equation, the equation of state of energy density can be given by
\begin{eqnarray}\label{eqq26}
{\omega _{de}} =  - 1 + \frac{2}{{3n}}\frac{{\sqrt {{\Omega _{de}}} }}{a}  .
\end{eqnarray}
Meanwhile using the Friedmann equations with equation Eq. (\ref{eqq26}), as well as ${\rho _m} \propto {a^{ - 3}}$, one can get ${\Omega _{de}}$ that satisfies \cite{Wei}
\begin{eqnarray}\label{eq27}
{\Omega '_{de}} = \frac{{{\Omega _{de}}}}{a}(1 - {\Omega _{de}})(3 - \frac{2}{n}\frac{{\sqrt {{\Omega _{de}}} }}{a})  ,
\end{eqnarray}
where the prime represents the derivative with respect to $\ln a$.

 It has some interesting features for the equation of state of the dark energy. In the dark energy dominated phase, the energy density can drive the universe to accelerated expansion if ${\omega _{de}} <  - 1/3$. From Eq. (\ref{eqq26}), it is easy to see that when $a \to \infty $, ${\Omega _{de}} \to 1$, thus ${\omega _{de}} \to  - 1$ at later time.  As well, in the matter dominated epoch, it has $a \propto {t^{2/3}}$, from Eq. (\ref{eqq23})(\ref{eqq24}), one has ${\rho _{de}} \propto {a^{ - 1}}$. Then from the dark energy density conservation equation, one can obtain ${\omega _{de}} =  - 2/3$.

  For constant ${\omega _{de}}$, the deceleration factor ${q_0}$ is given by ${q_0} = 0.5 + 1.5(1 - {\Omega _m}){\omega _{de}}$, we fix ${\Omega _m} = 0.3$ of the current universe which is from $\Lambda CDM$ cosmology with SNIa Pantheon samples \cite{Betoule}. It is easy to see that when ${\omega _{de}} \mathbin{\lower.3ex\hbox{$\buildrel<\over
{\smash{\scriptstyle\sim}\vphantom{_x}}$}}  - 1/2$, ${q_0} < 0$, which implies that the energy density can drive the universe to accelerated expansion if ${\omega _{de}} \mathbin{\lower.3ex\hbox{$\buildrel<\over{\smash{\scriptstyle\sim}\vphantom{_x}}$}}  - 1/2$ for the current universe.

\subsection{The invariable ground state energy for the particle
pairs and energy density}

If we also consider other case that the expected value of energy for the particle pair in the vacuum ${\varepsilon _c}$ does not change with time, ${\varepsilon _c}$ is constant, so Eq. (\ref{eq17}) becomes
\begin{eqnarray}\label{eq31}
u = {\varepsilon _c}\delta {t^{ - 3}}.
\end{eqnarray}

From Eq. (\ref{eq4}) and Eq. (\ref{eq31}), and introduce a constant ${n^2}$, we have
\begin{eqnarray}\label{eq32}
{\Omega _{de}} = \frac{{{n^2}}}{{t{\kern 1pt} {H^2}}}.
\end{eqnarray}

 Then we can use the SNIa Pantheon sample and the Planck 2018 CMB angular power spectra to constrain the parameters of dark energy models based on Eq. (\ref{eq25}) and (\ref{eq32}).

\section{The Used observation data}

\subsection{SNIa Pantheon sample and the Plank 2018 CMB angular power spectra   }
For SNIa data, the Pantheon sample is the combination of SNe Ia from the Pan-STARRS1 (PS1), the Sloan Digital Sky Survey (SDSS), SNLS, and various low-z and Hubble Space Telescope samples. The Panoramic Survey Telescope $\&$ Rapid Response System (Pan-STARRS or PS1) is a wide-field imaging facility built by the University of Hawaii's Institute for Astronomy, which is used for a variety of scientific studies from the nearby to the very distant Universe, and it has provided 279 SNe Ia for Pantheon sample \cite{Scolnic}. The Supernova Legacy Survey Program detected approximately 2000 high-redshift Supernovae between 2003 and 2008, and the Pantheon sample contains about 236 SNe Ia based on the first three years of data, which can be used to investigate the expansion history of the universe, and improve the constraint of cosmological parameters, as well as dark energy study \cite{Conley}.

 In 2014 SDSS Survey released a large catalog which contains light curves, spectra, classifications, and ancillary data of 10,258 variable and transient sources\cite{Gunn1998, York, Gunn2006, Sako2007, Betoule, Sako2014}. The release generated the largest sample of supernova candidates, and 500 samples have been confirmed as SNe Ia by the spectroscopic follow-up. 335 SNe Ia in the Pantheon sample is taken from this spectroscopic sample. The rest of the Pantheon sample are from the  CfA$1-4$, CSP, and Hubble Space Telescope (HST) SN surveys \cite{Conley}. This extended sample of 1048 SNe Ia is called the Pantheon sample.

 The Planck 2018 CMB angular power spectra data is based on observations obtained with Planck (http://www.esa.int/Planck), an ESA science mission with instruments and contributions directly funded by ESA Member States, NASA, and Canada.

\subsection{SALT2 calibration for Pantheon sample}
When correcting the apparent magnitude of the Pantheon sample, considering that the prior dark energy equation of state is unknown, so we use SALT2 and Taylor expansion of ${{\rm{d}}_H} - z$ relation to directly calibrate the distance modulus, which can simplify the problem.

The Taylor expansion of ${{\rm{d}}_H} - z$ relation can be given by
 \begin{equation}\label{eq36}
{{d_{H,th}} = \frac{1}{{1 - y}}\left\{ {y - \frac{{{q_0} - 1}}{2}{y^2} + \left[ {\frac{{3{q_0} - 2{q_0} - {j_0}}}{6} + \frac{{ - {\Omega _{{k_0}}} + 2}}{6}} \right]{y^3}} \right\}},
\end{equation}
where $y = z/(1 + z)$. In order to reduce calculation error for high redshift data, we take this variable substitution. ${q_0}$  is the deceleration parameter, ${j_0}$ is the jerk parameters, and ${\Omega _{{k_0}}}$ is the curvature term.

Meanwhile, the relation of distance modulus $\mu $ and luminosity distance ${d_H}$ can be written as
 \begin{eqnarray}\label{eq35}
\mu  = 5{\log _{10}}{d_H} + 25 - 5{\log _{10}}{H_0}.
\end{eqnarray}

We use SALT2 and Taylor expansion of ${{\rm{d}}_H} - z$ relation to directly calibrate Pantheon sample. The distance modulus ${\mu _{ob}}$ correction formula is given by SALT2 model \cite{Guy2005,Guy2007}
 \begin{eqnarray}\label{eq33}
{\mu _{B,ob}} = {m_B} - {M_B}{\rm{ + }}\alpha  \times {x_1}{\rm{ + }}\beta  \times c{\rm{ + }}\Delta B,
\end{eqnarray}
where ${m_B}$ corresponds to the observed peak magnitude in rest frame $B$ band, ${x_1}$ describes the time stretching of the light curve, $c$ describes the SN colour at maximum brightness, $\Delta B$ is a bias correction based on previous simulations, and$\alpha $, $\beta $ are nuisance parameters in the distance estimate. ${M_B}$ is the absolute B-band magnitude, which depends on the host galaxy properties \cite{Betoule}. Notice that ${M_B}$ is related to the host stellar mass (M stellar ) by a simple step function
 \begin{eqnarray}\label{eq34}
{M_B} = \left\{ \begin{array}{l}
M_B^1{\kern 1pt} {\kern 1pt} {\kern 1pt} {\kern 1pt} {\kern 1pt} {\kern 1pt} {\kern 1pt} {\kern 1pt} {\kern 1pt} {\kern 1pt} {\kern 1pt} {\kern 1pt} {\kern 1pt} {\kern 1pt} {\kern 1pt} {\kern 1pt} {\kern 1pt} {\kern 1pt} {\kern 1pt} {\kern 1pt} {\kern 1pt} {\kern 1pt} {\kern 1pt} {\kern 1pt} {\kern 1pt} {\kern 1pt} {\kern 1pt} {\kern 1pt} {\kern 1pt} {\kern 1pt} {\kern 1pt} {\kern 1pt} {\kern 1pt} {\kern 1pt} {\kern 1pt} if{\kern 1pt} {\kern 1pt} {\kern 1pt} {M_{stellar}} < {10^{10}}{M_ \odot }\\
M_B^1 + {\Delta _M}{\kern 1pt} {\kern 1pt} {\kern 1pt} {\kern 1pt} {\kern 1pt} {\kern 1pt} {\kern 1pt} {\kern 1pt} otherwise
\end{array} \right.
\end{eqnarray}
Here ${M_ \odot }$ is the mass of the Sun.

From Eq. (\ref{eq35}) and (\ref{eq33}), the ${\chi ^2}$ of Pantheon data can be calculated as
 \begin{eqnarray}\label{eq37}
{\chi ^2} = \Delta {\mu ^T}C_{{\mu _{ob}}}^{ - 1}\Delta \mu ,
\end{eqnarray}
where $\Delta \mu  = {\mu} - {\mu _{th}}$. ${C_\mu }$ is the covariance matrix of the distance modulus $\mu $, we only consider statistical error, and
 \begin{equation}\label{eq38}
\begin{array}{l}
{C_{\mu ,stat}} = {V_{{m_B}}} + {\alpha ^2}{V_{{x_1}}} + {\beta ^2}{V_c} + 2\alpha {V_{{m_B},{x_1}}} - 2\beta {V_{{m_{B,c}}}}\\
{\kern 1pt} {\kern 1pt} {\kern 1pt} {\kern 1pt} {\kern 1pt} {\kern 1pt} {\kern 1pt} {\kern 1pt} {\kern 1pt} {\kern 1pt} {\kern 1pt} {\kern 1pt} {\kern 1pt} {\kern 1pt} {\kern 1pt} {\kern 1pt} {\kern 1pt} {\kern 1pt} {\kern 1pt} {\kern 1pt} {\kern 1pt} {\kern 1pt} {\kern 1pt} {\kern 1pt} {\kern 1pt} {\kern 1pt} {\kern 1pt} {\kern 1pt} {\kern 1pt} {\kern 1pt}  - 2\alpha \beta {V_{{x_1},c}}
\end{array}
\end{equation}
From Eq. (\ref{eq37}), and in combination with the Pantheon sample, we obtain the statistical average and error for distance modulus $\mu $ and the parameters ${q_0}$, ${j_0}$. Then we use the calibrated Pantheon sample to constrain the dark energy models parameters.

\section{Using the Pantheon sample to fit the models parameters}
\subsection{The fitting of model  \uppercase\expandafter{\romannumeral1} parameters}

For model \uppercase\expandafter{\romannumeral1} ${\Omega _{de}} = {\raise0.7ex\hbox{${{n^2}{H^{ - 2}}}$} \!\mathord{\left/
 {\vphantom {{{n^2}{H^{ - 2}}} t}}\right.\kern-\nulldelimiterspace}
\!\lower0.7ex\hbox{$t$}}$, we firstly choose the universe age $T$ as the time scale $t$, and consider Taylor expansion of $T-z$ relation in nearly flat space, Which is

 \begin{eqnarray}\label{eq39}
T - {T_0} =  - \frac{1}{{{H_0}}}(y - \frac{{{q_0}}}{2}{y^2} + \frac{{2{q_0}^2 - {j_0}}}{6}{y^3}),
\end{eqnarray}
where ${\rm{y}} = z/(1 + z)$. In order to reduce calculation error for high redshift data, we take this variable substitution. ${T_0}$ is the present age of the universe.

Then the Hubble parameter in nearly flat space can be written as
 \begin{eqnarray}\label{eq40}
E(z) = \sqrt {{\Omega _m}{{(1 + z)}^3} + {\Omega _R}{{(1 + z)}^4} + \frac{{{n^2}}}{{T{\kern 1pt} {H_0}^2}}} .
\end{eqnarray}
where $E(z) = {\raise0.7ex\hbox{${H(z)}$} \!\mathord{\left/
 {\vphantom {{H(z)} {{H_0}}}}\right.\kern-\nulldelimiterspace}
\!\lower0.7ex\hbox{${{H_0}}$}}$. From Eq. (\ref{eq39}) and (\ref{eq40}), one has ${T_0} = {\raise0.7ex\hbox{${{n^2}{H_0}^{ - 2}}$} \!\mathord{\left/
 {\vphantom {{{n^2}{H_0}^{ - 2}} {(1 - {\Omega _m})}}}\right.\kern-\nulldelimiterspace}
\!\lower0.7ex\hbox{${(1 - {\Omega _m})}$}}$ when ${\Omega _R} \ll {\Omega _m}$.

 We use ${\chi ^2}$ statistic fitting method to constrain the parameters
  \begin{equation}\label{eq41}
{\chi ^2}_{Pantheon} = \Delta {\mu ^T}{C_\mu }^{ - 1}\Delta \mu  + \Delta {q_0}^2{\sigma _{{q_0}}}^{ - 2} + \Delta {j_0}^2{\sigma _{{j_0}}}^{ - 2},
 \end{equation}
where $\Delta \mu  = {\mu} - {\mu _{th}}$. $\Delta {q_0} = {q_0} - {q_{0,prior}}$ ,$\Delta {j_0} = {j_0} - {j_{0,prior}}$, ${q_{0,prior}},{j_{0,prior}}$ are given by SALT2 calibration method in Section 5.2.

\begin{figure*}[tbp]
\begin{center}
\includegraphics[scale=0.5]{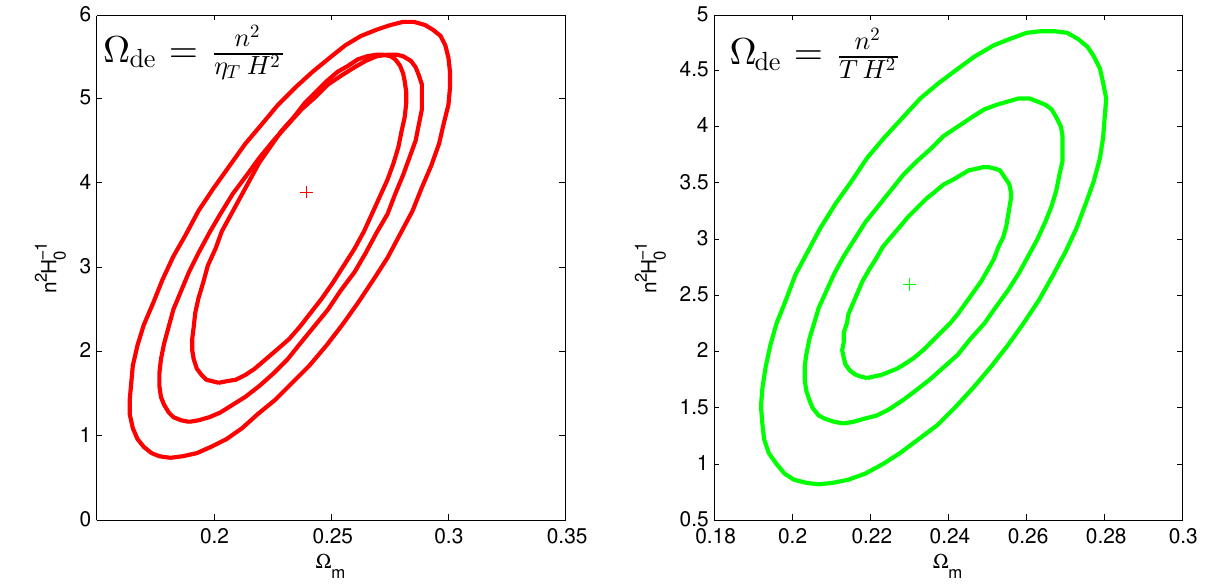}
  \caption{68.3\%, 95.4\%, and 99.7\% confidence region of the (${\Omega _m}$, ${n^2}{H_0}^{ - 1}$ ) plane from Pantheon observation data combined with Model \uppercase\expandafter{\romannumeral1}, the + dots in responding color represent the best fitting values for ${\Omega _m}$, ${n^2}{H_0}^{ - 1}$.}
  \label{fig1}
    \end{center}
\end{figure*}

Following this, we combine Pantheon data with Eq. (\ref{eq40}) to constrain the parameters,  and then use MCMC technology and ${\chi ^2}$ statistic fitting method to obtain the statistical mean values and the minimum chi-square without systematic errors of the parameters ${\Omega _m},{n^2}{H_0}^{ - 1},{q_0},{j_0}, {\kern 1pt} {t_0}{H_0}$, which are shown in Table (~\ref{tab1}), The confidence region of ${\kern 1pt} ({\Omega _m},{n^2}{H_0}^{ - 1})$ plane are 68.3\%, 95.4\% and 99.7\%. (see Figs. (~\ref{fig1}))

\begin{table*}
\footnotesize
 \centering
  \caption{ Statistical mean values of the cosmological parameters from SN Ia Pantheon sample observation data combined with Model \uppercase\expandafter{\romannumeral1}.}
  \label{tab1}
  \vspace{0.3cm}
  \begin{tabular}{@{}ccccccc@{}}
  \hline
   ${\Omega _{{\rm{de}}}} = {\raise0.7ex\hbox{${{n^2}{H^{ - 2}}}$} \!\mathord{\left/
 {\vphantom {{{n^2}{H^{ - 2}}} t}}\right.\kern-\nulldelimiterspace}
\!\lower0.7ex\hbox{$t$}}$&${\Omega _{\rm{m}}}$&${n^2}{H_0}^{ - 1}$&$q_0$&$j_0$&${\kern 1pt} {\kern 1pt} {t_0}{H_0}{\kern 1pt} ({t_0} = {\eta _{{T_0}}},{\kern 1pt} {\kern 1pt} {T_0}) $&$\chi _{\min }^2/d.o.f.$\\ \hline

   $t = {\eta _T}$&0.23$\pm$0.013 & 3.3$\pm$0.5   &   -0.57$\pm$0.28 & -0.2$\pm$2.7 &4.3$\pm$0.7 &1040$/$1050  \\
   $t = T$ &0.236$\pm$0.012 & 2.84$\pm$0.56   &   -0.44$\pm$0.31&-0.95$\pm$2.8 &3.7$\pm$0.74 &1044$/$1050  \\
  \hline
\end{tabular}
\end{table*}

Moreover, we can also choose the conformal age ${\eta _T}$ as the time scale $t$; and consider Taylor expansion of ${\eta _T} - z$ relation in near flat space, Which is

 \begin{equation}\label{eq42}
{\eta _T} - {\eta _{{T_0}}} = \frac{1}{{{H_0}}}\left\{ {y - \frac{{{q_0} - 1}}{2}{y^2} + \left[ {\frac{{3{q_0} - 2{q_0} - {j_0}}}{6}} \right]{y^3}} \right\},
\end{equation}
Where ${\eta _{{T_0}}}$ is the current conformal age of the universe.

The Hubble parameter in near flat space can be expressed as
 \begin{eqnarray}\label{eq43}
E(z) = \sqrt {{\Omega _m}{{(1 + z)}^3} + {\Omega _R}{{(1 + z)}^4} + \frac{{{n^2}}}{{{\eta _T}{\kern 1pt} {H_0}^2}}} .
\end{eqnarray}
Then from Eq. (\ref{eq42}) and (\ref{eq43}), ${\eta _{{T_0}}}$ satisfies
 ${\eta _{{T_0}}} = {\raise0.7ex\hbox{${{n^2}{H_0}^{ - 2}}$} \!\mathord{\left/
 {\vphantom {{{n^2}{H_0}^{ - 2}} {(1 - {\Omega _m})}}}\right.\kern-\nulldelimiterspace}
\!\lower0.7ex\hbox{${(1 - {\Omega _m})}$}}$.

In the same way, we use Eq. (\ref{eq43}) with the Pantheon sample to fit the model parameters, and the statistical results are shown in Table (\ref{tab1}), and Figs. (\ref{fig1}) shows the confidence region of ${\kern 1pt} ({\Omega _m},{n^2}{H_0}^{ - 1})$ plane are 68.3\%, 95.4\% and 99.7\%.

\subsection{The fitting of model \uppercase\expandafter{\romannumeral2} parameters}
When considering model \uppercase\expandafter{\romannumeral2}, in which ${\Omega _{de}} = {\raise0.7ex\hbox{${{n^2}{H^{ - 2}}}$} \!\mathord{\left/
 {\vphantom {{{n^2}{H^{ - 2}}} {{t^{ - 2}}}}}\right.\kern-\nulldelimiterspace}
\!\lower0.7ex\hbox{${{t^{ - 2}}}$}}$, we can also select the universe age $T$ as the time scale $t$, and the Hubble parameter in near flat space can be written as
 \begin{eqnarray}\label{eq44}
E(z) = \sqrt {{\Omega _m}{{(1 + z)}^3} + {\Omega _R}{{(1 + z)}^4} + \frac{{{n^2}}}{{T{{\kern 1pt} ^2}{H_0}^2}}} .
\end{eqnarray}

Alternatively, if  we select the conformal age ${\eta _T}$ as the time scale $t$, the Hubble parameter in near flat space  satisfies
 \begin{eqnarray}\label{eq45}
E(z) = \sqrt {{\Omega _m}{{(1 + z)}^3} + {\Omega _R}{{(1 + z)}^4} + \frac{{{n^2}}}{{{\eta _T}{{\kern 1pt} ^2}{H_0}^2}}} .
\end{eqnarray}

\begin{figure*}
\begin{center}
\includegraphics[scale=0.5]{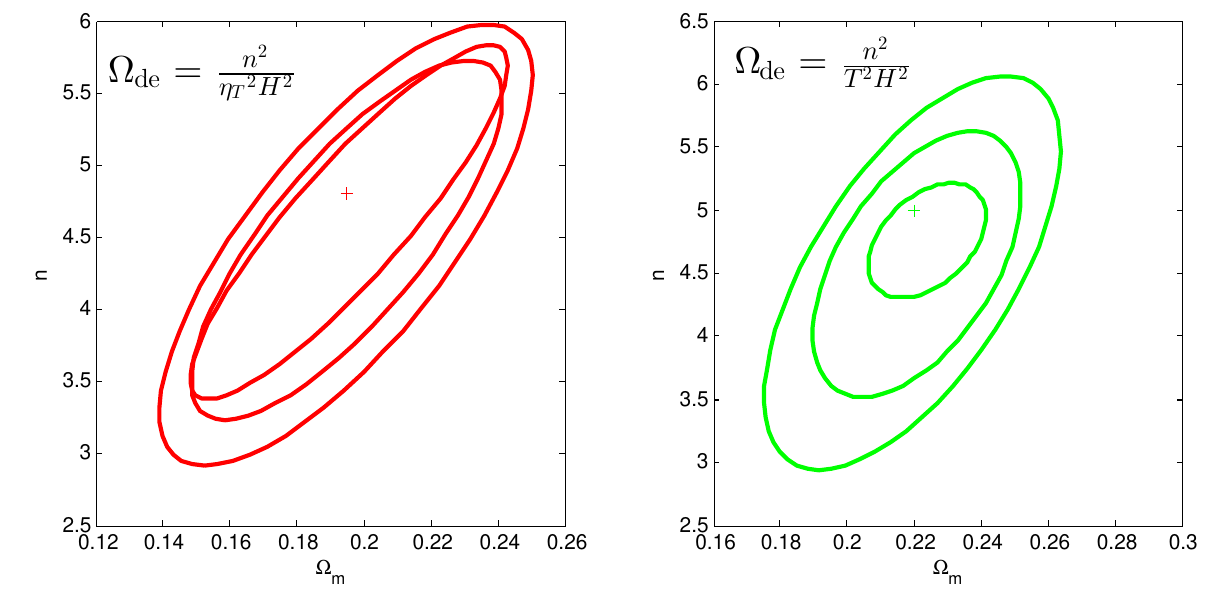}
  \caption{68.3\%, 95.4\%, and 99.7\% confidence region of the (${\Omega _m}$, $n$ ) plane from Pantheon observation data combined with Model \uppercase\expandafter{\romannumeral2}. The + dots in responding color represent the best fitting values for ${\Omega _m}$, $n$.}
  \label{fig2}
  \end{center}
\end{figure*}

Using the pointing models Eq. (\ref{eq44}) and (\ref{eq45}), we obtain the statistical results for the parameters which are shown in Table (\ref{tab2}),  and Figs. (\ref{fig2}) shows the confidence region of $({\Omega _m},n)$ plane are 68.3\%, 95.4\% and 99.7\%.

\begin{table*}
\footnotesize
 \centering
  \caption{ Statistical mean values of the cosmological parameters from SN Ia Pantheon sample observation data combined with Model \uppercase\expandafter{\romannumeral2}.}
  \label{tab2}
  \vspace{0.3cm}
  \begin{tabular}{@{}ccccccc@{}}
  \hline
   ${\Omega _{de}} = {\raise0.7ex\hbox{${{n^2}{H^{ - 2}}}$} \!\mathord{\left/
 {\vphantom {{{n^2}{H^{ - 2}}} {{t^2}}}}\right.\kern-\nulldelimiterspace}
\!\lower0.7ex\hbox{${{t^2}}$}}$&${\Omega _m}$&$n$&$q_0$&$j_0$&${\kern 1pt} {\kern 1pt} {t_0}{H_0}{\kern 1pt} ({t_0} = {\eta _{{T_0}}},{\kern 1pt} {\kern 1pt} {T_0}) $&$\chi _{\min }^2/d.o.f.$\\ \hline

   $t = {\eta _T}$&0.19$\pm$0.016 & 4.4$\pm$0.44   &   -0.64$\pm$0.27&-1.1$\pm$2.6 &5.5$\pm$0.62 &1042$/$1050  \\
   $t = T$ &0.22$\pm$0.01 & 4.5$\pm$0.4   &   -0.48$\pm$0.3&-0.7$\pm$2.8 &5.7$\pm$0.7 &1044$/$1050  \\
  \hline
\end{tabular}
\end{table*}

If we consider the current age of the universe to be ${T_0} = 13.5 \pm 0.5Gyr$ which is inferred from globular clusters \cite{Valcin}, and Hubble constant is ${H_0} = 73.5 \pm 1.4km{\kern 1pt} {\kern 1pt} {\kern 1pt} {\kern 1pt} {s^{ - 1}}{\kern 1pt} {\kern 1pt} {\kern 1pt} Mp{c^{ - 1}}$ using the NGC 4258 distance measurement \cite{Reid}. By combining with the fitting results obtained from the Pantheon sample in Table (\ref{tab1}) and (\ref{tab2}), we discover the choice of ${\eta _T}$ as time scale may be more persuasive. Furthermore, for the mean value of ${\Omega _m}$, Conley et al. provided a statistical result of matter density for constant $wCDM$ model ${\Omega _m} = 0.19_{ - 0.10}^{ + 0.08}$ from the combination of SNLS, HST, and low-z, SDSS data \cite{Conley},  our result is in agreement with it.

If only Pantheon data are used to investigate dark energy, the statistical results indicate the age dark energy models including ${\Omega _{de}} = {\raise0.7ex\hbox{${{n^2}{H^{ - 2}}}$} \!\mathord{\left/
 {\vphantom {{{n^2}{H^{ - 2}}} {{\eta _T}}}}\right.\kern-\nulldelimiterspace}
\!\lower0.7ex\hbox{${{\eta _T}}$}}$ and ${\Omega _{de}} = {\raise0.7ex\hbox{${{n^2}{H^{ - 2}}}$} \!\mathord{\left/
 {\vphantom {{{n^2}{H^{ - 2}}} {{\eta _T}^2}}}\right.\kern-\nulldelimiterspace}
\!\lower0.7ex\hbox{${{\eta _T}^2}$}}$ have no evident superiority compared to $\Lambda CDM$ using minimum chi-square.
\section{Using CMB angular power spectra to constrain the models parameters}
\begin{figure*}[htbp]
\begin{center}
\includegraphics[scale=0.7]{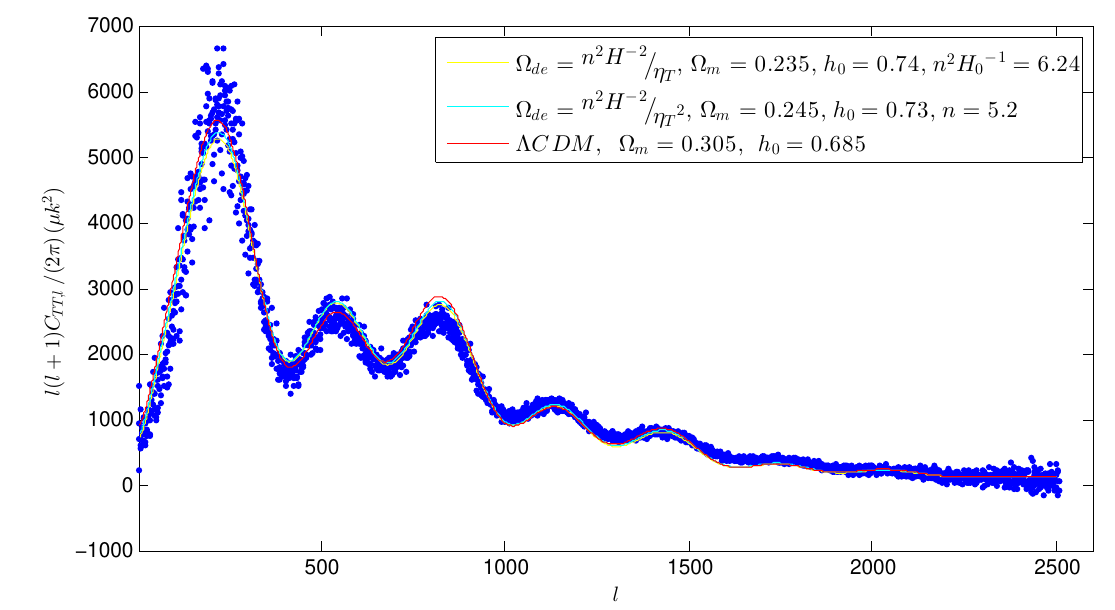}
  \caption{ Blue dots in responding color represent Planck 2018 CMB TT angular power spectra. The red, yellow and cyan lines are the CMB theoretical values of angular power spectra from $\Lambda CDM$, Model \uppercase\expandafter{\romannumeral1} and \uppercase\expandafter{\romannumeral2}, respectively, using the best fitting values, which are constrained by Planck 2018 data. }
  \label{fig3}
    \end{center}
\end{figure*}

In addition, we can use Planck 2018 CMB angular power spectra data to constrain the models parameters. When ${\rm{z}} \ge 2.5$, ${\raise0.7ex\hbox{${{\Omega _m}{{(1 + z)}^3}}$} \!\mathord{\left/
 {\vphantom {{{\Omega _m}{{(1 + z)}^3}} {{\Omega _{de}}(z)}}}\right.\kern-\nulldelimiterspace}
\!\lower0.7ex\hbox{${{\Omega _{de}}(z)}$}} \gg 1$, so we think the Taylor expansion for ${\eta _T} - z$ relation is also applicable to calculate CMB angular power spectra.
A calculation of $C_{TT,l}^s$ that ignores the  Sunyaev-Zeldovich (SZ) effect can refer to Weinberg 2008 \cite{Weinberg}, and SZ-effect can refer to Bond et al. 2005 \cite{Bond}. We use Planck 2018 data together with  Model \uppercase\expandafter{\romannumeral1} and \uppercase\expandafter{\romannumeral2} to constrain the cosmology parameters, and the statistical mean values of the parameters are shown in Table (\ref{tab3}), and the Figs. (\ref{fig3}) shows the CMB theoretical TT power spectra by the best fitting values from the Planck 2018.

In Table (\ref{tab3}), we provide the statistical mean values of Hubble constant ${H_0}{\rm{ = }}73.2 \pm 1.3{\rm{km}}{\kern 1pt} {\kern 1pt} {\kern 1pt} {s^{ - 1}}{\kern 1pt} {\kern 1pt} Mp{c^{ - 1}}$, which is consistent with the result obtained using NGC 4258 distance measurement \cite{Reid}.

For the dark energy density, when using Planck 2018 CMB data, the statistical results indicate the age dark energy models including ${\Omega _{de}} = {\raise0.7ex\hbox{${{n^2}{H^{ - 2}}}$} \!\mathord{\left/
 {\vphantom {{{n^2}{H^{ - 2}}} {{\eta _T}}}}\right.\kern-\nulldelimiterspace}
\!\lower0.7ex\hbox{${{\eta _T}}$}}$ and ${\Omega _{de}} = {\raise0.7ex\hbox{${{n^2}{H^{ - 2}}}$} \!\mathord{\left/
 {\vphantom {{{n^2}{H^{ - 2}}} {{\eta _T}^2}}}\right.\kern-\nulldelimiterspace}
\!\lower0.7ex\hbox{${{\eta _T}^2}$}}$ have evident superiority, compared to $\Lambda CDM$ using a minimum chi-square of $l(l + 1)C_{TT,l}^s{\kern 1pt} (l = 30 \sim 1500)$ .

\begin{table*}
\normalsize
 \centering
    \caption{The statistical mean values of the cosmological parameters from The Planck 2018 TT power spectra data combined with the Model \uppercase\expandafter{\romannumeral1} and \uppercase\expandafter{\romannumeral2}.}
 \label{tab3}
  \begin{tabular}{@{}ccc@{}}
  \hline
  &${\Omega _{de}} = {\raise0.7ex\hbox{${{n^2}{H^{ - 2}}}$} \!\mathord{\left/
 {\vphantom {{{n^2}{H^{ - 2}}} {{\eta _T}}}}\right.\kern-\nulldelimiterspace}
\!\lower0.7ex\hbox{${{\eta _T}}$}}$&${\Omega _{de}} = {\raise0.7ex\hbox{${{n^2}{H^{ - 2}}}$} \!\mathord{\left/
 {\vphantom {{{n^2}{H^{ - 2}}} {{\eta _T}^2}}}\right.\kern-\nulldelimiterspace}
\!\lower0.7ex\hbox{${{\eta _T}^2}$}}$ \\ \hline

   ${\Omega _b}{h^2}$&0.0214$\pm$0.00012&0.023$\pm$0.0002\\
   ${\Omega _c}{h^2}$&0.11$\pm$0.0014& 0.107$\pm$0.0014\\
   ${10^{10}}{A_s}{e^{ - 2\tau }}$&   1.29$\pm$0.016&   1.316$\pm$0.027\\
   ${n_s}$ &0.93$\pm$0.004&0.97$\pm$0.0025\\
   ${N_{eff}}$&3.45$\pm$0.13&3.5$\pm$0.35\\
   ${h_0}$ &0.742$\pm$0.012&0.732$\pm$0.013\\
   ${\Omega _m}$&0.237$\pm$0.018&0.243$\pm$0.022\\
   ${n^2}{H_0}^{ - 1},{\kern 1pt} {\kern 1pt} {\kern 1pt} n $&5.5$\pm$0.33&4.85$\pm$0.54\\
   $q_0$  &-0.38$\pm$0.24&-0.7$\pm$0.59\\
   $j_0$&1.3$\pm$0.22 &4.7$\pm$1.2 \\
   $\sigma _8^{sz}$&1.1$\pm$0.011 &1.06$\pm$0.012 \\  \hline

\end{tabular}
\end{table*}

\section{CONCLUSIONS}

 Understanding the physical nature of dark energy is important for our universe. In addition to the study of particle physics,  dark energy may also enable us to further explore the nature of the vacuum. Whether or not the dark energy is derived from particle
pairs, another particle, or neither,  still needs to be further verified.

We explore a theoretical possibility that dark energy density is derived from massless particle pairs in a vacuum. Assuming massless particle
pairs fall into the horizon boundary with the expansion of the universe, the quantum fluctuation of space-time enables us to estimate dark energy density. Meanwhile, to explain the physical nature of negative pressure, this is deduced from the increasing entropy density with the rise in the energy level $k$ in the Boltzmann system.

Next, we used the SNIa Pantheon sample and Planck 2018 CMB angular power spectra to constrain the specified models. The statistical results indicate that age dark energy models have evident superiority, when only using CMB data compared to $\Lambda CDM$ using minimum chi-square. Furthermore, we obtain the statistical mean value of the Hubble constant ${H_0}{\rm{ = }}73.2 \pm 1.3{\rm{km}}{\kern 1pt} {\kern 1pt} {\kern 1pt} {s^{ - 1}}{\kern 1pt} {\kern 1pt} Mp{c^{ - 1}}$, which is consistent with the result obtained using NGC 4258 distance measurement.

Finally, we extend our discussion to the future of the universe. From Eq. (\ref{eq22}), if ${k}$ satisfies $k:1 \to 2 \to 3 \cdot  \cdot  \cdot  \to {k_c}$, the universe will change from accelerating expansion to decelerating. Thus, the property of dark energy will dominate the future of the universe,  and it will similarly determine the future of humanity.

\section{Acknowledgments}
  This work was supported by Xiaofeng Yang{'}s Xinjiang Tianchi Bairen project and CAS Pioneer Hundred Talents Program. This work was also partly supported by the National Key R\&D Program of China under Grant No.2018YFA0404602.


\clearpage


\end{document}